# FLATTENING OF GALAXIES OF DIFFERENT MORPHOLOGICAL TYPES IN SUBCLUSTERS OF Coma

N. G. Kogoshvili,[1] T. M. Borchkhadze,[1] and A. T. Kalloghlian[2]

*This paper deals with the observed variation in the flattening of galaxies with the density of galaxies in the subclusters of Coma surrounding NGC 4889, NGC 4874, and NGC 4839 based on data from the Abastumani Merged Catalog of Galaxies. The mean values of the observed ratios of the diameters of the galaxies, as well as histograms of their distributions, indicate that in the central, dense regions of the subclusters within a volume of $0.5 h_{75}^{-1}$ Mpc³, E and S0 type galaxies are close to spheroidal. A significant reduction in the mean values of the diameters of the galaxies in the subclusters is noted, regardless of their morphology relative to the galaxies in the halo of the Coma cluster. In the subclusters, spiral galaxies are found with a hydrogen deficit that is more than 5 times the hydrogen deficit in spirals within the halo of the cluster. According to their 3-D coordinates, most of the galaxies with a hydrogen deficit are located closer to the south-east edge of the subcluster surrounding NGC 4874 near an extended gas filament in the x-ray region. This may indicate that the subcluster is moving toward a central condensation of faint galaxies in the Coma cluster and a possible merger with it.*

Keywords: *galaxies: clusters: subclusters*

## 1. Introduction

This is an examination of the effect of the local density of galaxies and the intergalactic medium in the Coma cluster on the formation of structure of the galaxies.

Based on a study of 55 rich clusters of galaxies, Dressler [1] concluded that there is a close correlation between the morphological type of a galaxy and the local density of galaxies. He found that bright galaxies tend to group preferentially in a high density region as opposed to faint galaxies which are more uniformly distributed.

In a model of the collision of galaxies in rich clusters, Richstone [2] discovered the structural changes and estimated the loss of matter and the reduction in radius of the galaxies in the central dense region of clusters over times comparable to the age of the universe. According to Richstone [3], if elliptical galaxies are formed as a result of the interaction of disk-shaped systems, this may lead to the discovery of spherically symmetric structures in the dense central region of clusters similar to the Coma cluster.

By studying the surface brightness profiles of E and S0 galaxies in 6 rich clusters, including Coma, Strom and Strom [4-6] found that E galaxies in the dense central regions of clusters are smaller in size than the E galaxies observed in the halo of the clusters. At the same time, they believe that the ellipticity of E galaxies is independent of their position in a cluster and no tendency toward spherically symmetric shapes was observed in bright E galaxies.

In an analysis of the changes in the morphology of galaxies within dense regions of clusters based on modelling the rapid passage of disk galaxies through the central regions of clusters, Farouki and Shapiro [7] assumed that the excess of E galaxies in the center of clusters is caused by tidal interactions of disk galaxies leading to distortion of their disk.

In a study of the ellipticity of galaxies of different kinds in regions with high and low densities based on an examination of data of about 7000 galaxies in the UGC Nilson catalog [9], De Freitas Pacheco et al. [8] noted that the S0 and S objects are flatter in regions with a high density of galaxies.

Using Dressler's data [11] on galactic clusters and relying on models of the distribution of galaxies in clusters, Quintana et al. [10] found that in most clusters the segregation of galaxies on morphological type and luminosity correlates with the density of galaxies in subclusters that have not yet reached the relaxation stage.

## 2. Data and analysis

Up to now the frequency distribution of the ellipticities (flatness) of galaxies has been examined in dense central regions of individual rich clusters as compared with their extended, less dense halos.

In this paper we study the variation in the observed flatness of galaxies as a function of the local density of galaxies in subclusters selected in the Coma cluster by Kogoshvili et al. [12] using data from the Abastumani Merged Catalogue of Galaxies compiled by Kogoshvili and Borchkhadze [13]. Based on the method of hierarchical clustering and taking into account gravitational interaction among galaxies the 6 subclusters have been selected in [12], from which two central subclusters around NGC 4889 and NGC 4874 and one subcluster in the SW condensation of galaxies around NGC 4839 were considered as containing sufficient number of galaxies for analysis.

The distribution of the observed ratio of the diameters of the galaxies was examined as a function of the spatial density of galaxies in two regions of the subclusters:

(1) in the interior, denser regions of the subclusters, which are predominantly populated by E and S0 galaxies;

(2) in the outer, less dense regions of the subclusters with a mixed population, but with a predominance in the number of spiral galaxies; and, additionally,

(3) in the halo of the Coma cluster.

The samples studied here contain elliptical, lenticular, and spiral galaxies brighter than magnitude 15.5 with measured radial velocities, for which values of the logarithm of the angular diameter of the galaxy $\log D_{25}$ and the logarithm of the ratio of the angular diameters $\log R_{25} = \log(D/d)_{25}$, as well as the morphological type, are given in the RCG3 catalog of De Vaucouleurs et al. [14]

TABLE 1. Mean Values of Observed Quantities: Radial Velocity $\langle V \rangle$, Logarithm of the Ratio of the Diameters $\langle \log R_{25} \rangle$, Estimates of the Ratio of the Axes $\langle d/D \rangle$, and Absolute Magnitude $\langle M \rangle$ for E, S0, and S Galaxies in Subclusters Surrounding the Galaxies NGC 4889, 4874, and 4839

|  |  | $\langle V \rangle$ | E | | | S0 | | | S | | |
|---|---|---|---|---|---|---|---|---|---|---|---|
|  |  |  | $\langle \log R_{25} \rangle$ | $\langle d/D \rangle$ | $\langle M \rangle$ | $\langle \log R_{25} \rangle$ | $\langle d/D \rangle$ | $\langle M \rangle$ | $\langle \log R_{25} \rangle$ | $\langle d/D \rangle$ | $\langle M \rangle$ |
| Subcluster I with NGC 4889 |  | 6522 | 0.05 | 0.89 | -19.5 | 0.19 | 0.65 | -19.6 | 0.28 | 0.52 | -19.8 |
|  | σ | 703 | 0.09 |  | 0.5 | 0.18 |  | 0.6 | 0.24 |  | 0.7 |
|  | n | 33 | 9 |  | 13 | 7 |  | 7 | 9 |  | 12 |
| Subcluster II with NGC 4874 |  | 7038 | 0.04 | 0.91 | -19.9 | 0.09 | 0.81 | -19.7 | 0.33 | 0.47 | -20.0 |
|  | σ | 636 | 0.03 |  | 0.3 | 0.08 |  | 0.3 | 0.20 |  | 0.8 |
|  | n | 35 | 3 |  | 6 | 9 |  | 10 | 15 |  | 19 |
| Subcluster III with NGC 4839 |  | 7254 | 0.04 | 0.91 | -20.2 | 0.10 | 0.79 | -19.9 | 0.43 | 0.37 | -19.6 |
|  | σ | 425 | 0.05 |  | 0.4 | 0.11 |  | 1.0 | 0.27 |  | 0.5 |
|  | n | 42 | 9 |  | 2 | 10 |  | 11 | 12 |  | 16 |

An analysis of the observed ellipticities of the galaxies depends significantly on the reliability of the data that are used and, in particular, on the reliability with which the morphological type has been determined and on the assumptions made about the random orientation of the galaxies.

According to Table 1, the small sample for the ratio of the galactic diameters, mainly of early types, based on data from RCG3, is a consequence of a lack of measurements of the diameters of the galaxies which do not have distinct boundaries within the limits of a given isophote. The mean values of the observed ellipticities of the galaxies tend toward spherical shapes in the case of E galaxies over the entire volume of the three subclusters. This tendency is more noticeable in subcluster I surrounding NGC 4889, where E galaxies constitute a majority in the central region. As for the S0 galaxies, which form a compact group around NGC 4874 in the central region of subcluster II, they have a mean observed ellipticity of $\langle d/D \rangle = 0.8$, as do the S0 objects in subcluster III, which also indicate their proximity to spherical shape.

In the course of analyzing the observed flatnesses of the galaxies, the estimated accuracy of the isophote measurements was taken into account: on the order of 10% for the galactic diameters and a more precise value $0.2 \div 0.4\%$ for the ratio of the galactic diameters obtained by Thompson [15] based on the data of 8 clusters, including Coma.

For a comparative analysis of the ellipticity of the galaxies observed in the denser central regions of the subclusters and in their less dense outer regions, as well as in order to enhance the statistics for galaxies of every morphological type, the data from the three subclusters were combined into a single sample.

Table 2 lists the calculations for the galaxies in the inner, denser region of the subclusters within a volume of $\sim 0.5 h_{75}^{-1}$ Mpc$^3$ (panel I) and in the outer region within a volume of $\sim 0.5 - 1 h_{75}^{-1}$ Mpc$^3$ (panel II). Panel III lists the calculations for the galaxies in the halo of the Coma cluster.

TABLE 2. Mean Values of Observed Quantities: $\langle\log R_{25}\rangle$, $\langle d/D\rangle$, $\langle\log D_{25}\rangle$, and $\langle M\rangle$ for E, S0, and S Galaxies Based on Combining the Calculations of Three Subclusters Surrounding NGC 4889, 4874, and 4839

| | E | | | | S0 | | | | S | | | |
|---|---|---|---|---|---|---|---|---|---|---|---|---|
| | $\langle\log R_{25}\rangle$ | $\langle d/D\rangle$ | $\langle\log D_{25}\rangle$ | $\langle M\rangle$ | $\langle\log R_{25}\rangle$ | $\langle d/D\rangle$ | $\langle\log D_{25}\rangle$ | $\langle M\rangle$ | $\langle\log R_{25}\rangle$ | $\langle d/D\rangle$ | $\langle\log D_{25}\rangle$ | $\langle M\rangle$ |
| | Inner region of the subclusters | | | | | | | | | | | |
| | 0.04 | 0.91 | 0.87 | -19.7 | 0.09 | 0.81 | 0.87 | -19.8 | 0.41 | 0.39 | 0.97 | -19.6 |
| σ | 0.07 | | 0.22 | 0.5 | 0.10 | | 0.17 | 0.5 | 0.24 | | 0.13 | 0.5 |
| n | 15 | | 15 | 21 | 16 | | 16 | 17 | 16 | | 16 | 19 |
| | Outer region of the subclusters | | | | | | | | | | | |
| | 0.07 | 0.85 | 1.01 | -20.2 | 0.17 | 0.68 | 1.00 | -19.7 | 0.30 | 0.50 | 1.04 | -20.0 |
| σ | 0.05 | | 0.24 | 0.4 | 0.16 | | 0.18 | 1.0 | 0.23 | | 0.19 | 0.8 |
| n | 6 | | 6 | 9 | 10 | | 11 | 11 | 20 | | 20 | 28 |
| | Galaxies in the halo of the Coma cluster | | | | | | | | | | | |
| | 0.10 | 0.79 | 1.05 | -20.0 | 0.18 | 0.66 | 1.03 | -19.7 | 0.41 | 0.39 | 1.05 | -19.7 |
| σ | 0.07 | | 0.22 | 1.0 | 0.19 | | 0.17 | 0.5 | 0.26 | | 0.15 | 0.6 |
| n | 8 | | 8 | 16 | 14 | | 14 | 16 | 45 | | 45 | 57 |

The mean values $\langle\log R_{25}\rangle$ given for the E galaxies in Table 2 and the histograms of the distributions of $\log R_{25}$ according to Fig. 1 indicate that in the inner (panel I) and outer (panel II) parts of the subclusters, E galaxies are characterized by ellipticities within the range $\varepsilon = 10(1-b/a) = 0 \div 2.5$ with an estimate for the number of E0 galaxies on the order of 70%, which suggests that the elliptical galaxies in the subclusters are close to spherical in shape. For the S0 galaxies in the inner parts of the subclusters, there is less tendency toward spheroidal shapes, with ellipticities in the range $\varepsilon = 0 \div 3.5$, while in their less dense outer regions there is a substantially larger number of more flattened S0 objects.

Unlike the distribution of the ratio of diameters for early galaxy types which has a distinct maximum around shapes that are close to spheroidal, the spiral galaxies in the dense inner region of the subclusters are characterized by two peaks in the distribution of the ellipticity around $\varepsilon = 2$ and 6, and in the outer parts of the subclusters, a concentration of the ellipticities around two distinct maxima at $\varepsilon = 1.5$ and 5. In the inner regions of the two central subclusters, there is a small number of spiral galaxies.

There is a slight tendency for the ellipticity of the E galaxies to increase in the less dense parts of the subclusters; this is more noticeable in the case of the S0 galaxies, and increases for the galaxies in the field of the Coma cluster. At the same time, an increase is seen in the mean diameter of the galaxies, regardless of their morphology on going from the subclusters to the field of the Coma cluster.

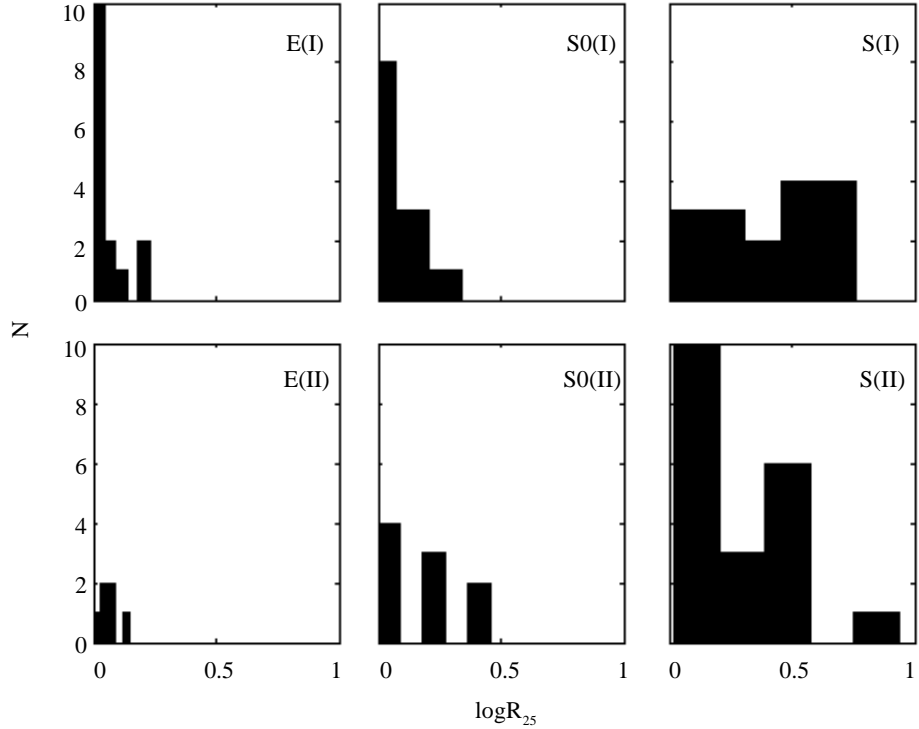

Fig. 1. Histograms of the distribution of observed values of $\log R_{25}$ for E, S0, and S galaxies from Table 2 for the inner (row I, top) and outer (row II, bottom) regions of the subclusters.

## 3. Comparative analysis of results

The structural changes in the galaxies may depend on the local density of galaxies and, thereby, on the tidal interaction between galaxies, their rates of collision, and interactions with intercluster medium.

According to Richstone [2], in regions of rich clusters where the density of galaxies is high, gravitational interactions take place often enough when galaxies lose stars from weakly bound galaxy halos and thereby are reduced in luminosity and radius. Richstone estimates that the reductions in absolute magnitude and radius of E galaxies are on the order of $\Delta M \sim 0^m.4$ and $\Delta \log r \sim 0.15$. The corresponding estimates by Strom and Strom [4] are $\Delta M \sim 0^m.5$ and $\Delta r_{26} \sim 0.09$.

In the selected subclusters we have received the characteristic average decrease in the brightness and diameter of the E galaxies on going from the inner, dense regions to the less dense regions of the subclusters as $\Delta M \sim 0^m5$ and $\Delta \log D \sim 0.14$. On the other hand, we estimate an overall reduction of $\Delta \log D \sim 0.18$ in the diameter of the E galaxies on going from the dense regions of the subclusters to the halo of the Coma cluster.

Richstone [2], Farouki and Shapiro [7], and Dekel et al. [16] point out that the interaction of galaxies in the dense medium may cause formation of galaxies with spherically symmetric shapes. Although this effect is small, it may show up through a change in the ratio of the diameters of the galaxies in regions with different densities.

In the case of the subclusters, the average reduction in the ellipticity of the E galaxies is by $\Delta d/D \sim 0.06$ on going from the denser to the less dense regions, and by $\Delta d/D \sim 0.12$ on going from the dense regions of the subclusters to the halo of the Coma cluster.

Based on measurements of the ratio of the diameters of E galaxies in a central region of the Coma cluster of size $r < 1$ Mpc, Strom and Strom [4] give a ratio of $N_{\varepsilon<2}/N_{\varepsilon>3} = 2.39$ for E galaxies with different ellipticity without finding a significant dependence of the ellipticity of the E galaxies on their location within the cluster.

Contrary to the tendency of elliptical and spiral galaxies to have spherical shapes in dense regions of clusters, De Freitas Pacheco et al. [8]. Dekel et al. [16], and De Souza et al. [17] observed the opposite effect for S0 galaxies, specifically, flatter shapes of S0 galaxies in these regions.

Our estimates in Table 2 show that the S0 galaxies in the denser, interior regions of the subclusters have nearly spherical shapes and are characterized by higher luminosities than the S0 galaxies observed in the outer regions and in the halo of the Coma cluster. The reduction in the ratio of the axes of these galaxies is $\Delta d/D \sim 0.13$ on the average on going from the inner to the outer portions of the subclusters, while the corresponding change is $\Delta d/D \sim 0.15$ on going from the dense regions of the subclusters to the halo of the cluster. At the same time, according to Table 1, the tendency to spherical shapes is manifested by S0 galaxies primarily in the denser regions of subclusters II and III, which may also be indicative of a greater bulge in these galaxies. Dressler [18], in turn, has pointed out that the S0 galaxies in rich clusters are predominantly systems with large bulges.

As for the spiral galaxies, De Freitas Pacheco et al. [8] note a significant concentration in the observed flattening of the spiral galaxies around two peaks, 0.15 and 0.55, both in the general field of the galaxies and in the regions with an elevated density of galaxies. This is consistent with our calculations only in the case of the spiral galaxies in the subclusters.

The difference between our results and those previously published can be explained in a number of cases by a different approach to selecting the galaxies for structural studies in the Coma cluster. In particular, the dependence on the spatial density of galaxies in the subclusters was compared with the dependence of the visible density of galaxies in a cluster; when the galaxies in the foreground and background of the halo of the cluster, projected onto its central regions, could distort the calculated results.

## 4. The hydrogen deficit in spiral galaxies

Dekel et al. [16] have pointed out that spiral galaxies in regions with a high density of galaxies often have a gas deficit and that spiral galaxies which are rich in gas are essentially never encountered in these regions.

Dressler [18] reached a similar conclusion based on an analysis of a catalog of spiral galaxies with a hydrogen HI deficit compiled by Giovanelli and Hayes [19].

In an earlier paper [12] we detected galaxies with a hydrogen deficit in subclusters of Coma in terms of the parameter $HI = m_{21}^0 - B_T^0$ given in the RCG3 catalog. Based on observations of spiral galaxies in the 21 cm line in the region of the Coma supercluster to find galaxies with a hydrogen deficit, Gavazzi et al. [20] introduced the parameter DefHI, which was defined taking the optical linear diameter of the galaxies into account. We have used data from the catalog of Gavazzi in order to greatly increase the number of galaxies with an HI deficit in the sample of spiral galaxies.

The comparison of the estimated HI deficit of the spiral galaxies according to the two catalogs in Fig. 2 shows a fair agreement between the determinations of the parameter DefHI by the two methods for $n = 37$ comparison galaxies and corresponds to a correlation coefficient $r = 0.87$.

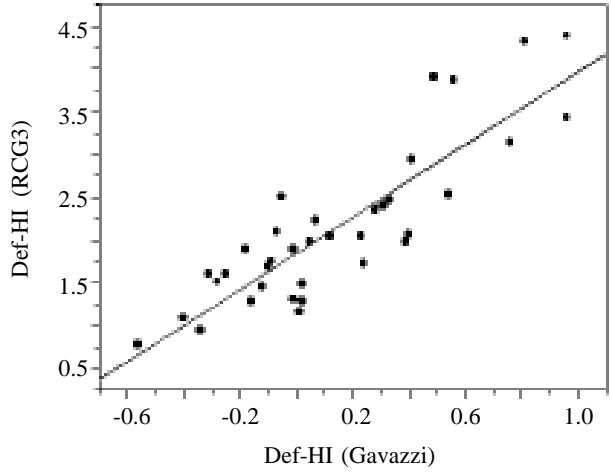

Fig. 2. The relationship between the parameters Def-HI determined from RCG3 and the catalog of Gavazzi et al. [20]

Table 3 shows that the spiral galaxies with a hydrogen deficit in the subclusters and in the halo have similar mean values of the ratio of the diameters but with larger hydrogen deficit and smaller angular sizes of the galaxies in the subclusters. For the galaxies which do not manifest a hydrogen deficit in the subclusters, flatter structures are seen, but with more flattening and lower luminosity of the spiral galaxies in the cluster halo.

An analysis of Table 4 reveals greater flattening and a greater HI deficit for the spiral galaxies, mostly of the early subtypes Sa and Sb according to the Gavazzi catalog [20], in the dense inner regions of the subclusters compared to the spiral galaxies in their outer, less dense regions. At the same time, the mean values of the ratio of the diameters are the same for spiral galaxies, regardless of the hydrogen HI content in the inner region, and for spirals without an HI deficit in the outer regions of the subclusters.

TABLE 3. Mean Values of $<$Def-HI$_{(Gavazzi)}>$, $<\log R_{25}>$, $<\log D_{25}>$ and $<M>$ for Spiral Galaxies in the Subclusters and in the General Field of the Coma Cluster

|   | | Galaxies with HI deficit | | | Galaxies without HI deficit | | |
|---|---|---|---|---|---|---|---|
|   | Def-HI$_{Gavazzi}$ | $<\log R_{25}>$ | $<\log D_{25}>$ | $<M>$ | $<\log R_{25}>$ | $<\log D_{25}>$ | $<M>$ |
|   | Galaxies in the subclusters in Coma | | | | | | |
|   | 0.54 | 0.31 | 1.00 | -19.9 | 0.42 | 1.02 | -19.7 |
| σ | 0.44 | 0.23 | 0.19 | 0.7 | 0.24 | 0.12 | 0.7 |
| n | 22 | 24 | 24 | 26 | 12 | 12 | 20 |
|   | Galaxies in the halo of the Coma cluster | | | | | | |
|   | -0.01 | 0.34 | 1.07 | -19.9 | 0.51 | 1.02 | -19.5 |
| σ | 0.30 | 0.21 | 0.18 | 0.5 | 0.31 | 0.10 | 0.6 |
| n | 26 | 26 | 26 | 28 | 19 | 19 | 29 |

TABLE 4. Mean Values of <Def-HI$_{(Gavazzi)}$>, <log$R_{25}$> and <M> for the Spiral Galaxies Based on Combined Calculations in the Inner and Outer Regions of the Coma Subclusters

|  | Galaxies with deficit | | | | Galaxies without deficit | | | |
|---|---|---|---|---|---|---|---|---|
|  | <Def-HI$_{(Gavazzi)}$> | <log$R_{25}$> | <d/D> | <$D_{25}$> | <M> | <log$R_{25}$> | <d/D> | <$D_{25}$> | <M> |
| Inner region of the subclusters | | | | | | | | | |
|  | 0.66 | 0.41 | 0.39 | 0.98 | -19.8 | 0.43 | 0.37 | 0.93 | -19.6 |
| σ | 0.33 | 0.25 |  | 0.14 | 0.4 | 0.23 |  | 0.10 | 0.5 |
| n | 12 | 11 |  | 11 | 12 | 5 |  | 5 | 5 |
| Outer region of the subclusters | | | | | | | | | |
|  | 0.44 | 0.23 | 0.63 | 1.01 | -20.1 | 0.42 | 0.38 | 1.06 | -19.9 |
| σ | 0.45 | 0.18 |  | 0.22 | 0.9 | 0.25 |  | 0.10 | 0.7 |
| n | 15 | 13 |  | 13 | 15 | 7 |  | 7 | 13 |

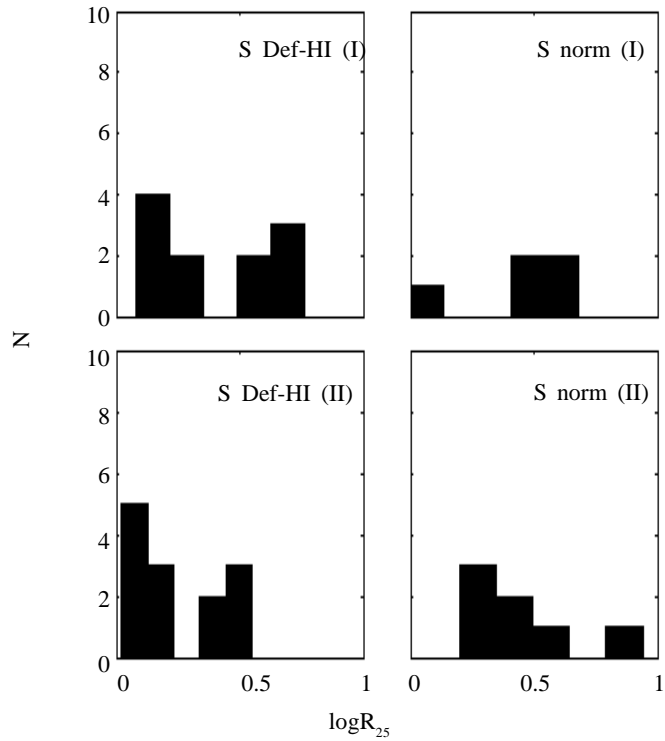

Fig. 3. Histograms of the distribution of observed values of log$R_{25}$ for spiral galaxies with and without hydrogen HI deficit in the inner (row I) and outer (row II) regions of the subclusters.

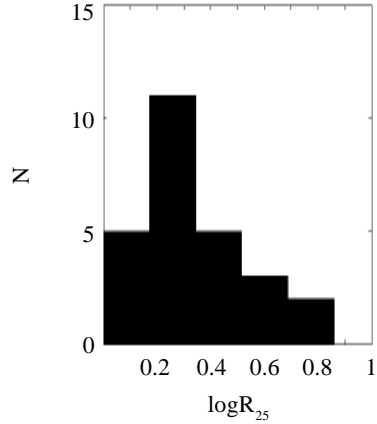

Fig. 4. Histogram of the distribution of observed values of log$R_{25}$ for spiral galaxies with a hydrogen HI deficit in the halo of the Coma cluster.

Figure 3 shows histograms of the distribution of log$R_{25}$ for spiral galaxies with an HI deficit in the inner (top) and outer (bottom) regions of the subclusters. The concentration of galaxies around two peaks with estimated ellipticities of $\varepsilon = 0 \div 2$ and 5-6 is noticeable, with a large number of galaxies having images close to circular and a smaller number of flatter structures. The small number of spiral galaxies in dense inner regions of the subclusters suggests, according to Merritt [21], that these galaxies have elongated radial orbits and do not spend a significant time in these regions. The smaller diameters and luminosities of the spiral galaxies with a hydrogen deficit in the denser regions of the subclusters might, as can be seen from Table 4, be the result of a change in the structure of the galaxies owing to loss of gas and stars from the halo of the galaxies as they interact in the dense medium.

Figure 4 shows that the spiral galaxies with an HI deficit in the field of the Coma cluster have only a single maximum in the distribution at log$R_{25} = 0.25$.

## 5. Discussion of results

Of the 11 spiral galaxies in the dense inner region of the subclusters, 5 are mentioned in Bravo-Alfaro's list [22] as objects with the most perturbed hydrogen HI structure. Of these, 4 lie in the central region of the subcluster surrounding NGC 4874 and 2 coincide with radio galaxies from the list of Venturi et al. [23]. Of the 18 spiral galaxies in this subcluster, 10 have a hydrogen deficit with an average value of <Def-HI$_{Gavazzi}$> = 0.51 ± 0.38.

Vikhlinin et al. [24] have discovered a ~ 1$h_{50}^{-1}$ Mpc long gas filament in the X-ray region, lying to the south-east of the center of the Coma cluster. They explain the formation of the filament in terms of the removal of gas from the group of galaxies owing to their interaction with the intergalactic medium. Donnelly et al. [25] believe that this group of galaxies is associated with NGC 4874 and is moving toward the center of the Coma cluster, followed by the gas filament. We believe that this group of galaxies is the spiral galaxies with Def-HI that we have selected in the subcluster surrounding NGC 4874. This is confirmed by the histograms of the distributions with respect to $\alpha$, $\delta$, and $V$ of the Def-HI galaxies in Fig. 5 (top) as compared with analogous histograms for the distributions of the other galaxies in the subcluster surrounding NGC 4874 (bottom). According to Fig. 5, in terms of $\alpha$ and $\delta$ most of the Def-HI galaxies lie near the south-east edge of the main group of galaxies in the subcluster, and their velocities (>7000 km/s) may indicate that they are most likely on the far from the observer side of the subcluster.

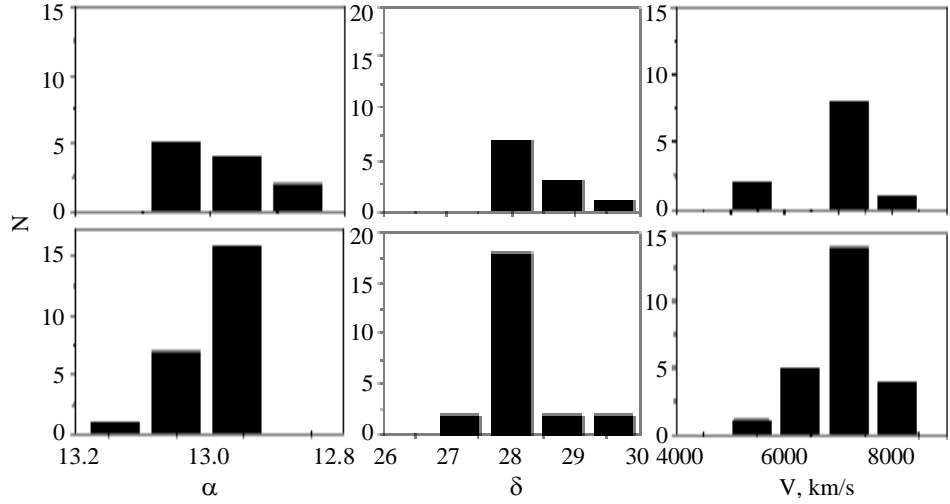

Fig. 5. Histograms of the distribution of galaxies with Def-HI in the subcluster surrounding NGC 4874 with respect to $\alpha$, $\delta$, and V (row I) and for the remaining galaxies in the same subcluster (row II).

The formation of a central dominant galaxy in clusters is usually related to a distribution of surrounding faint galaxies with a high density and a high probability of collisions among them. In particular, Merrifield et al. [26] have analyzed the distribution of galaxies brighter than $20^m$ within 250 $h^{-1}$ kpc of the central galaxy using CCD images of 29 galactic clusters and found that they subsequently merge with the central object. Based on data from the Godwin [28] catalog of galaxies brighter than $b_{26.5} = 20^m$, Baier et al. [27] found an excess in the distribution of faint galaxies near NGC 4874 as the only central dominant galaxy of the Coma cluster. On the other hand, in an analysis of the structure of the central region of the Coma cluster based on data from the Godwin catalog, Biviano et al. [29] found a smooth distribution of faint galaxies with a maximum between the central galaxies NGC 4874 and NGC 4889 of the subclusters that was slightly shifted toward NGC 4874. In the opinion of Biviano, a merger of the two central subclusters surrounding NGC 4874 and NGC 4889 with the main body of the cluster is being observed, which indicates that the Coma cluster is in the process of formation as it merges with groups of galaxies from surrounding structures.

## 6. Conclusions

Calculated mean values of the ratio of the diameters of E and S0 galaxies and histograms of their distributions in two central subclusters of Coma surrounding NGC 4889 and NGC 4874, as well as in the SW condensation surrounding NGC 4839, indicate that in the inner, most dense regions of the subclusters within a volume of $0.5 h_{75}^{-1}$ Mpc$^3$, the E and S0 galaxies are close to spheroidal in shape. At the same time, the flatness of these galaxies tends to increase on going to less dense regions of the subclusters within a volume of $0.5 \div 1 h_{75}^{-1}$ Mpc3 and beyond that to the halo of the Coma cluster. The mean angular diameter of the galaxies is found to increase significantly, regardless of their morphological type, in the outer, less dense regions of the subclusters and it increases considerably in the halo of the cluster. There are two maxima in the observed flatness distribution of spiral galaxies over the entire volume of the subclusters with a large number of galaxies having images that are close to circular and a smaller number of more flattened structures.

The dependence of the flatness of the galaxies on their density in the subclusters can be explained by tidal interactions between the galaxies in these subclusters, such that stellar clusters and stars that have been removed from the outer regions of the galaxies by collisions can merge with the massive central galaxy and cause its brightness to increase with time.

Objects with a hydrogen deficit roughly a factor of 5 times greater than that in galaxies within the field of the Coma cluster have been discovered among the spiral galaxies in the subclusters that we have isolated. According to their 3-D coordinates, most of the galaxies with a hydrogen deficit are closer to the south-east boundary of the subcluster surrounding NGC 4874 near an extended gas filament (in the X-ray region) which has most likely been formed by the ejection of gas from these galaxies as they interact with the hot intergalactic medium; this may indicate that the subclusters are moving toward and merging with the central condensation of faint galaxies in the Coma cluster.